\documentclass[twocolumn,preprintnumbers,amsmath,amssymb]{revtex4}
\usepackage{graphicx}% Include figure files
\usepackage{dcolumn}% Align table columns on decimal point
\usepackage{bm}% bold math
\begin{document}
\title{\bf Effect of grain boundary on the buckling of graphene nanoribbons}
\author{M. Neek-Amal$^{1}$ and F. M. Peeters$^2$ }
\affiliation{$^1$Department of Physics, Shahid Rajaee University,
Lavizan, Tehran 16785-136, Iran.\\$^2$Departement Fysica,
Universiteit Antwerpen, Groenenborgerlaan 171, B-2020 Antwerpen,
 Belgium.}
\date{\today}

\begin{abstract}
The buckling of graphene nano-ribbons containing a grain boundary are
studied using atomistic simulations where free and supported
boundary conditions are invoked.  We found that when
graphene contains a small angle grain boundary the buckling strains are larger when the ribbons with
free  (supported) boundary condition are subjected to compressive
 tension parallel (perpendicular) to the grain boundary. The shape of the deformations of
the buckled graphene nanoribbons depends  on the boundary
conditions and the presence of the grain boundary and the direction
of applied in-plane compressive tension. Large angle grain boundary
results in smaller buckling strains as compared to
perfect graphene or to a small angle grain boundary.
\end{abstract}
\maketitle

 Large area graphene sheets have been grown on metallic foils which were found
to contain grain boundaries~\cite{nature2004}. Scanning tunneling
microscopy experiments were used to investigate  tilt grain
boundaries in graphite~\cite{surfacescience}. Transmission electron
microscopy was able to observe individual dislocations in
graphene~\cite{science324}. The effect of grain boundaries on both
electronic and mechanical properties of graphene was investigated
theoretically by Yazyev \emph{et al}~\cite{prb81}. Depending on the
grain boundary structure, high transparency and perfect reflection
of charge carriers over remarkably large energy ranges was reported
by using first principle calculations~\cite{nature2010}.
The effects of grain boundaries on the buckling
of nano-scale graphene nanoribbons (GNR) has not been investigated up to now, while it is important
 for their mechanical stability.

Recently, we studied the effect of  applied external axial stress on the thermomechanical properties
of perfect graphene (PG)~\cite{prb82neek,JPCMneek} and GNRS containing randomly distributed
vacancies~\cite{APLneek}. In this paper we address the effect of the presence of grain
boundaries and in-plane boundary compressive stress applied in two different directions, on the buckling
and the stability of GNR for the case of supported boundary (SBC) and free boundary
conditions (FBC).
We found
that the presence of grain boundaries: 1) alters the sine wave shape
of the longitudinal deformation
 modes of GNR when subjected to SBC,
 2) results in a concave shape of the GNRs when subject to FBC, 3)
when subject to compressive tension perpendicular  to the grain boundary, the GNR  buckles at smaller (larger)
strains as compared to perfect graphene in case of FBC (SBC), 4)
the buckling transition found for compressive tension parallel to
the grain boundary is three times larger than the one for perfect graphene subjected to compressive
tension along the zig-zag direction, and 5) free energy calculations reveal that a
larger angle grain boundary in suspended graphene when it is subjected to   compressive
 tension makes it unstable.

\begin{figure}
\begin{center}
\includegraphics[width=0.52\linewidth]{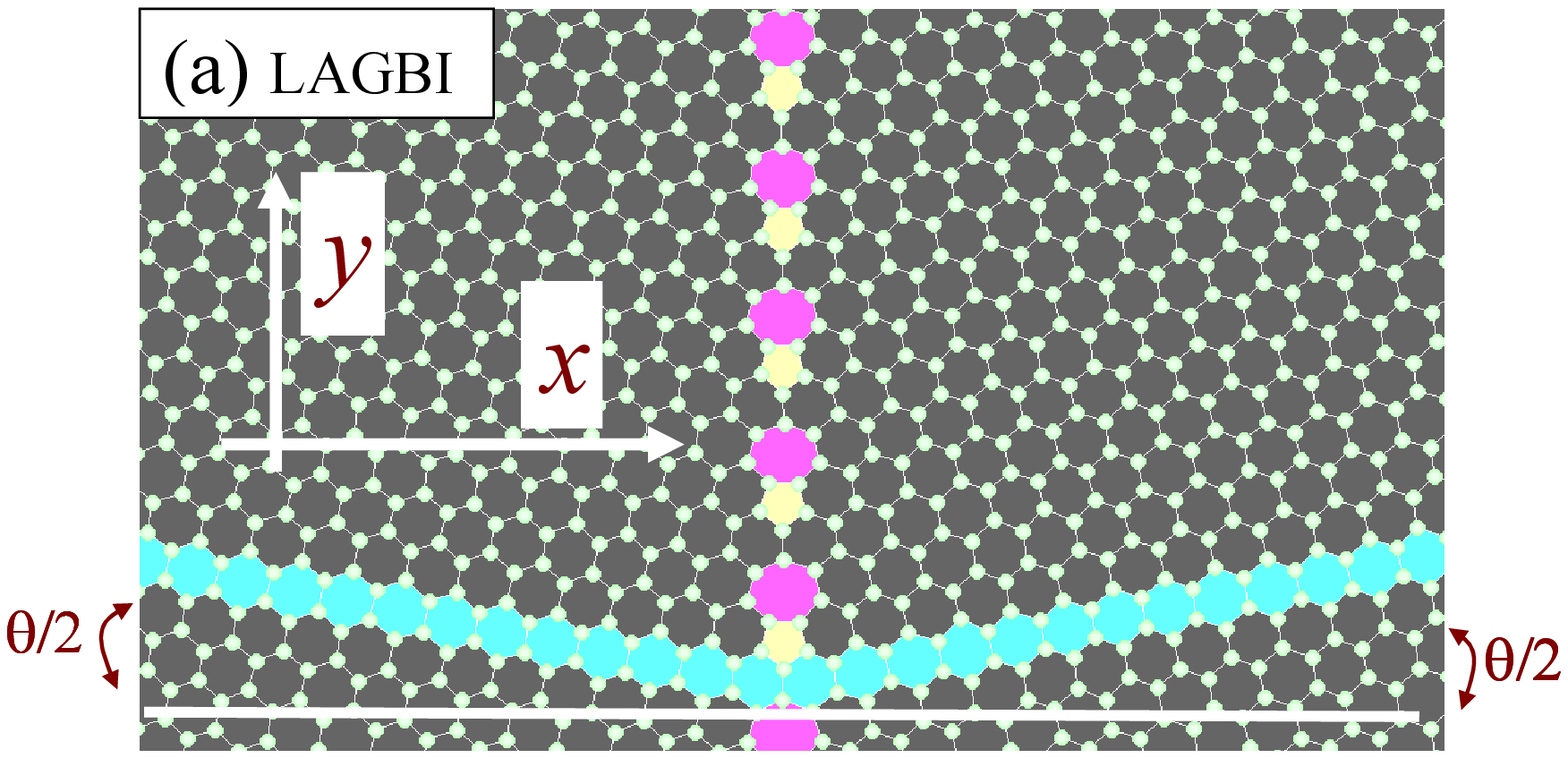}
\includegraphics[width=0.465\linewidth]{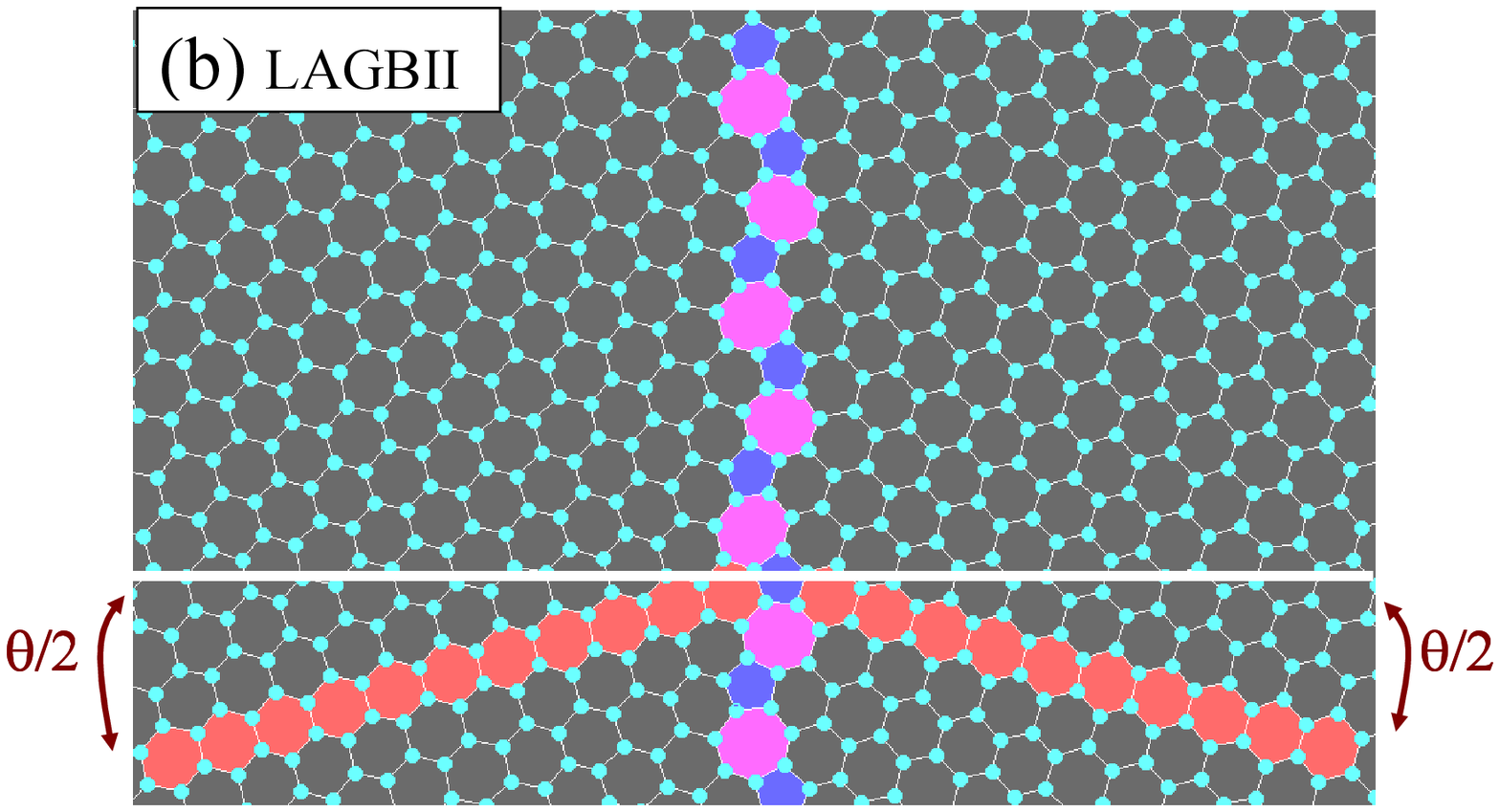}
 \caption{(Color
online) Two snap shots taken from the central portion of GNRs with
grain boundaries  after relaxation at room temperature. Two
different configurations  of the grain boundaries (LAGBI in (a) with
$\theta=$21.8$^o$ and LAGBII in (b) with $\theta=$32.2$^o$, see
Ref.~\cite{prb81}) are shown by the colored pentagon and heptagons .
\label{figmodel} }
\end{center}
\end{figure}

Initially the coordinates of all atoms are put in a flat surface of
a honey-comb lattice with nearest neighbor distance equal to
$a_0=0.142~nm$. Our perfect GNR (PG) is a rectangular GNR with
dimensions $a\times b= 20\times 10$\,nm$^2$ in $x$ and $y$ directions with armchair (ac)
and zigzag (zz) edges, respectively. A grain boundary is introduced
as an array of 5-7 defects, which are put in the center of the PG
along the y-direction with angle $\theta$ (see
Figs.~\ref{figmodel}). \textbf{As an example,  we study two kinds of grain boundaries
 which were named  LAGBI and LAGBII by Yazyev \emph{et al}~\cite{prb81}.
These grain boundaries are typical interfaces between
domains of graphene with different crystallographic orientation.
Mutual orientations of the two crystalline domains is described by
the misorientation angle which for LAGBI is $\theta=$21.8$^o$ and for LAGBII is $\theta=$32.2$^o$.
 %The number of atoms in PG is 15\,960 and in LAGBI is 15\,510.
In Figs.~\ref{figmodel} we depict two
snap shots of the central portion of
  LAGBI (a) and LAGBII (b) after relaxation at room temperature.}

  Classical atomistic molecular dynamics simulations (MD) are employed to
simulate compressed PG,  LAGBI and LAGBII using Brenner's bond-order
potential~\cite{brenner} and  temperature is controlled by a
Nose'-Hoover thermostat at room temperature.
%\textbf{ We cut the edges of the defected graphenes sheet such that they have  a rectangular
%shape with dimension $20\times 10$\,nm$^2$ (the main system) however the perfect sheet can be
%originally created in rectangular shape and no cut is necessary.
%}
Before starting the compression, the systems are equilibrated during
75 ps (150\,000 time steps).
%\textbf{ We used the following method for applying the compression and boundary conditions:
%a few
\textbf{ Extra atoms were added to the boundaries of the rectangular samples
which are characterized by:
$x=\pm a/2\pm~ 2\AA$,~ $y=\pm b/2\pm~ 2\AA$. The compression and boundary conditions (FBC and SBC) are applied
 on these extra atoms which are outside the main systems.
 % with
%$a\times b= 20\times 10\,nm{^2}$ dimensions. For example when we compress the system a
%long $x$-direction the boundary atoms at $|x|>a/2$~ (we named longitudinal ends) are used for
%applying the compression and boundary atoms  at $|y|>b/2$~(we named lateral ends)
%are used for applying the boundary conditions and vise versa for applying compression along $y$-direction.
%In fact the main system (inside the boundary atoms) receives all deformations from these boundary atoms.
Compressing direction is defined by the angle `$\alpha$'. \textbf{For example $\alpha=0~ (\pi/2)$
implies that  compression
is applied in the $x$-direction ($y$-direction)  so that
 the right (up) longitudinal ends at $x>a/2$ ($y>b/2$) are under compressive tension in $-x$($-y$)-direction and  left longitudinal ends at $x<-a/2$ ($y<-b/2$) are under  compressive tension
  in $x$ ($y$)-direction
   (see Fig. 1 in Ref.~\cite{prb82neek}).
Note that only in PG  `ac'(`zz') direction is equivalent to $x$-direction ($y$-direction).
 For the lateral ends (for $\alpha=0$ they are $|y|>b/2$ and for $\alpha=\pi/2$ they are $|x|>a/2$),  we used
SBC, (only movement of atoms in lateral ends in the compression direction is allowed not in the $z-$direction) and FBC.
The FBC (SBC) condition is equivalent to  suspended graphene at two longitudinal ends that is put over a trench
 while it is free at the lateral two ends~\cite{meyer}.
% Free boundary condition for our suspended graphene is related to the experimentally
% suspended graphene~\cite{meyer} from two longitudinal ends and
Supported boundary condition can be  created in experiments by suspending graphene from two longitudinal
ends and supporting the other two lateral ends to a substrate which prevents graphene to move verticaly~\cite{naturenanotechnology}.}
%By decreasing the coordinates of atoms in longitudinal ends where
% $x>a/2$ (when $\alpha$=0) or $y>b/2$ (when $\alpha=\pi/2$) and   $x<-a/2$ (when $\alpha$=0) or $y<-b/2$ (
% when $\alpha=\pi/2$) gradually with
}

\textbf{We applied a strain rate $\nu=0.027$/ns  and $\nu=0.054$/ns  for $\alpha=0$ and  $\alpha=\pi/2$, respectively.
The strain rate is given by $\nu= 2\delta x/5000 l \delta t$, where $\delta t=0.5$\,fs is our
 MD-simulation time step, $\delta x=0.667$\,pm is used as the compression step after
each 5000 steps and the factor two is because the compression is applied on two longitudinal  ends (for more details see
Ref.~\cite{prb82neek}). Here $l=a$ or  $l=b$ if $\alpha=0$ or $\alpha=\pi/2$, respectively.
 Notice that the atoms in the longitudinal ends are fixed during each compression step.}

\begin{figure*}
\begin{center}
\includegraphics[width=0.323\linewidth]{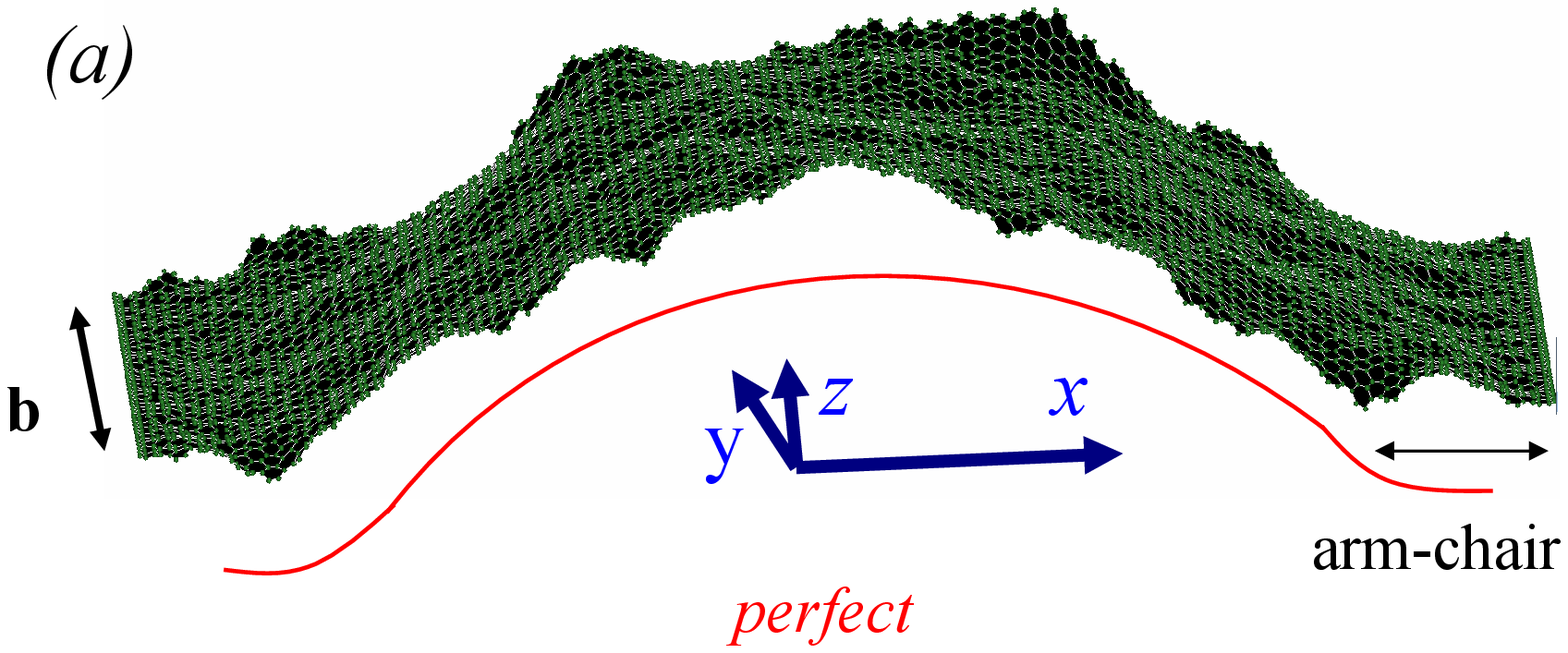}
\includegraphics[width=0.323\linewidth]{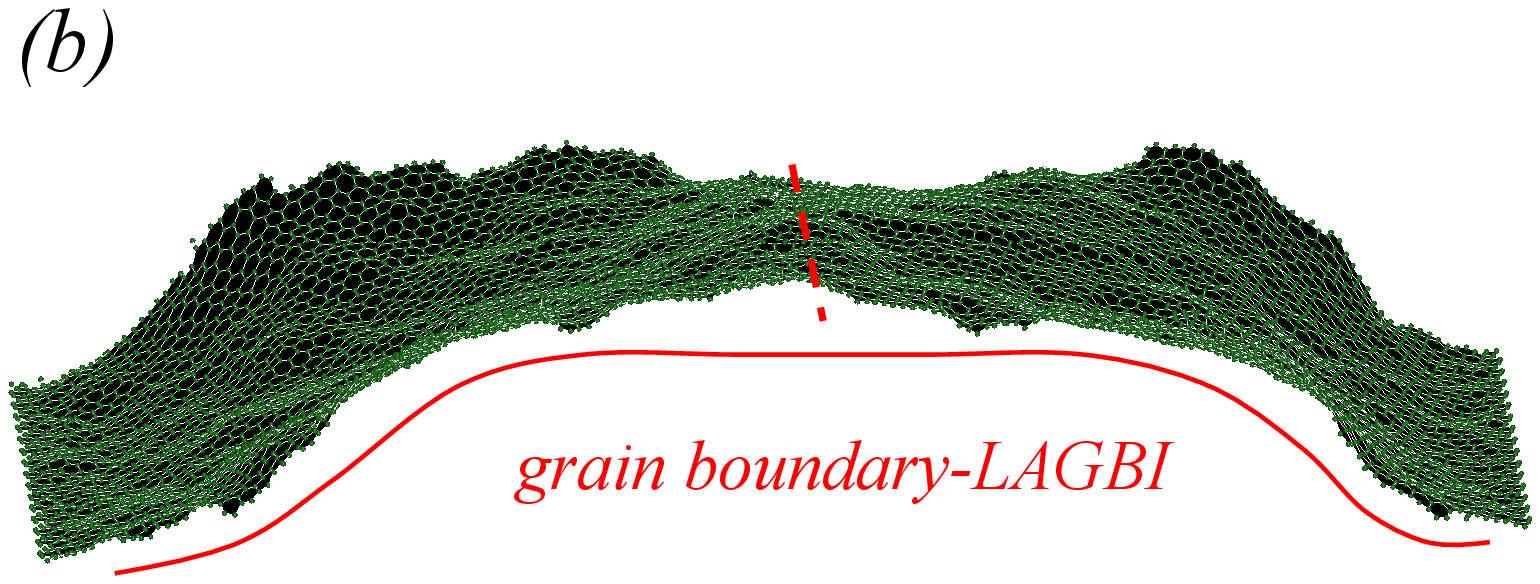}
\includegraphics[width=0.323\linewidth]{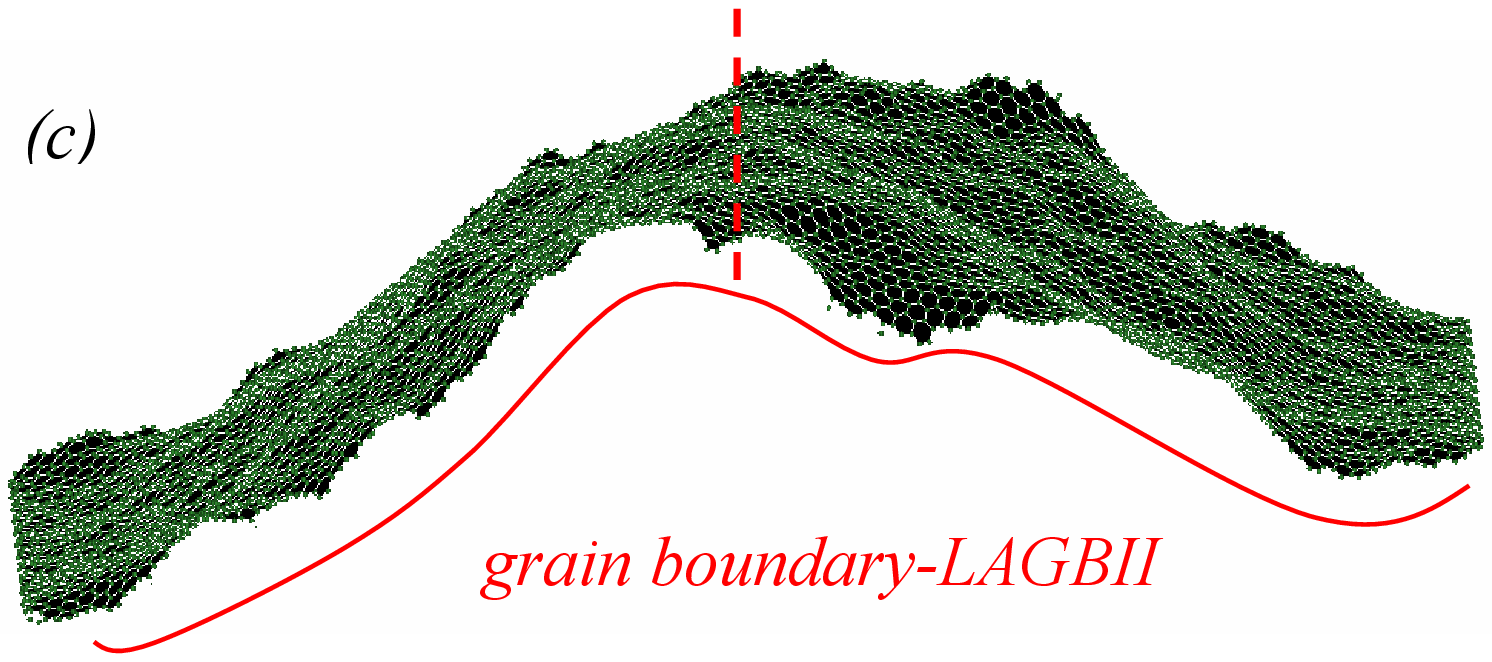}
\includegraphics[width=0.323\linewidth]{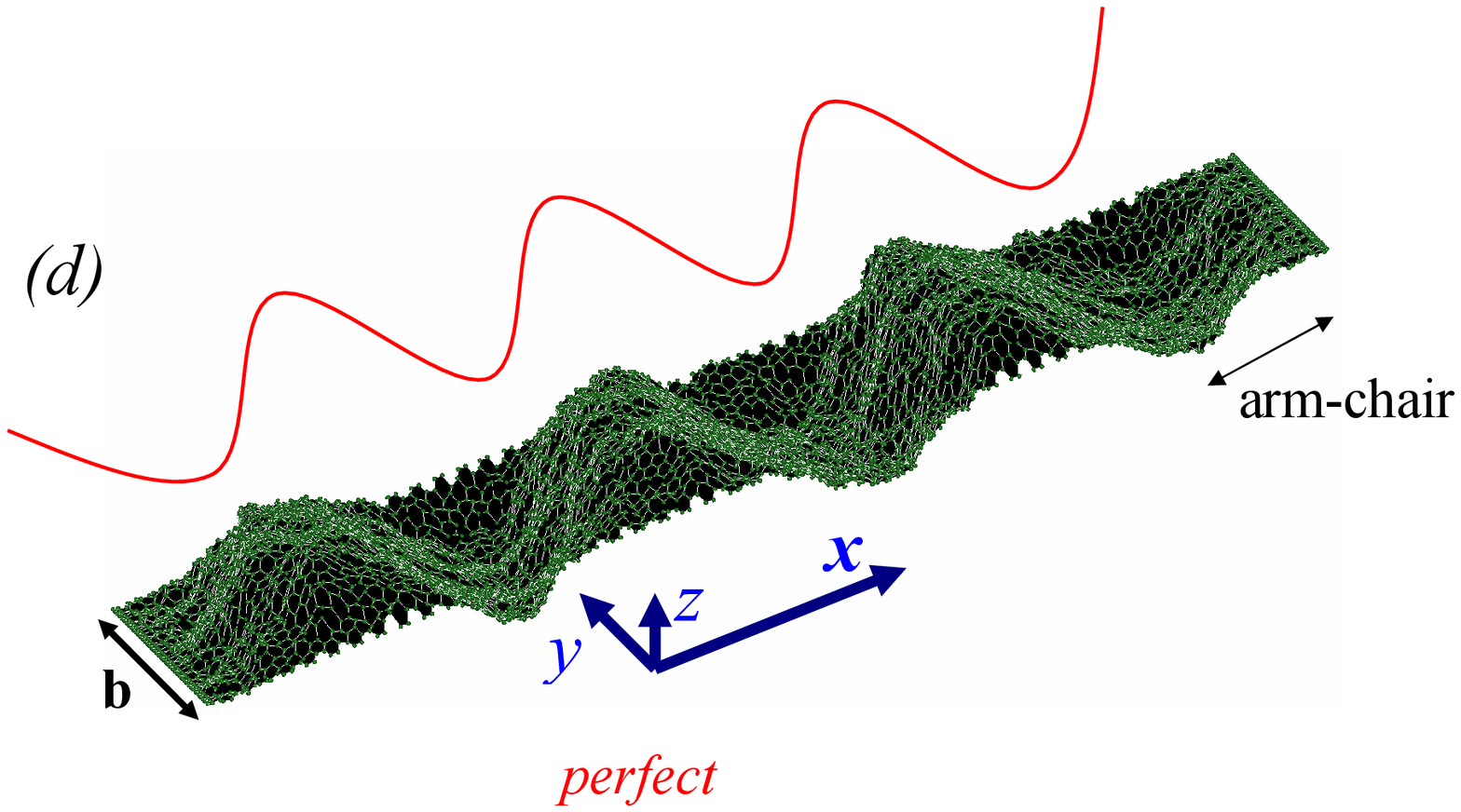}
\includegraphics[width=0.323\linewidth]{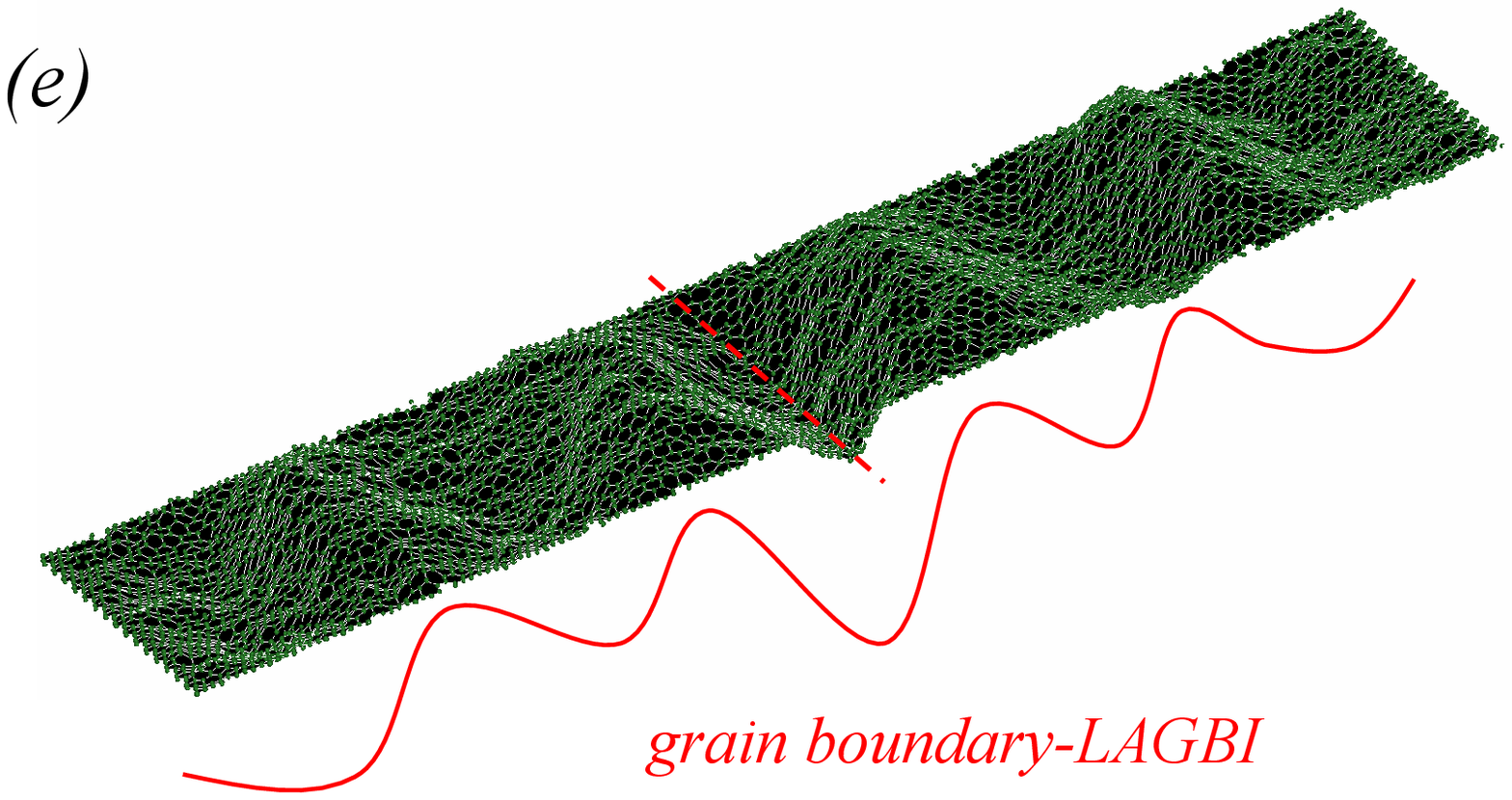}
\includegraphics[width=0.323\linewidth]{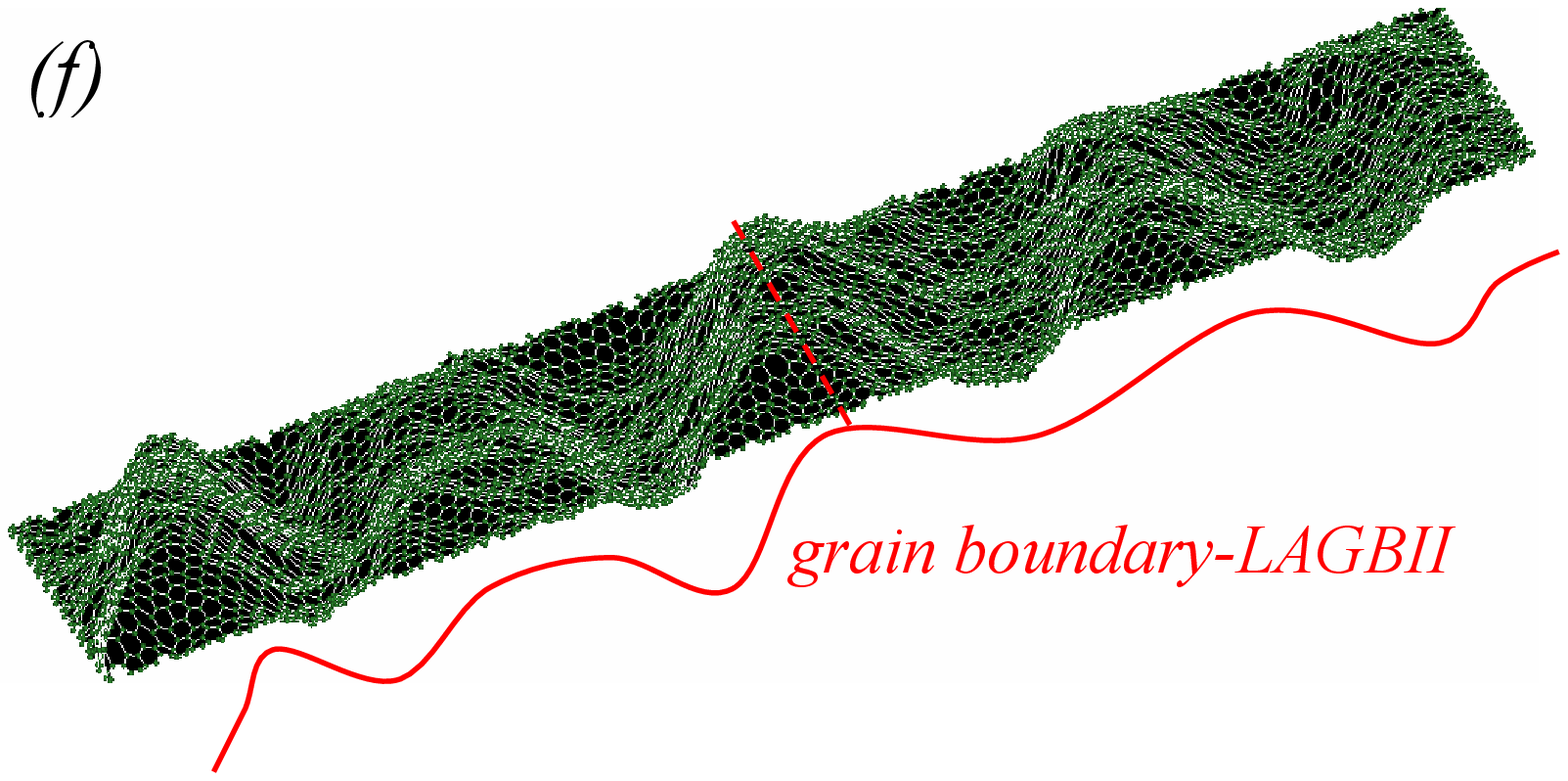}
\caption{(Color online) \textbf{Snap shots of  buckled graphene with
free lateral boundary condition (a,b,c) and supported lateral
boundary condition (d,e,f)
 for perfect graphene (a,d), graphene with LAGBI type grain boundary (b,e) and graphene with LAGBII type
  grain boundary (c,f) where $\epsilon=2.45\%$, $\alpha=0$. The red
curves give the $z$-deviation averaged over the $y$-direction (to improve visualization the $z$-components
 of all atoms were scaled by a
factor of 3 and the edge atoms were excluded)}. \label{fig2} }
\end{center}
\end{figure*}

\begin{figure*}
\begin{center}
\includegraphics[width=0.25\linewidth]{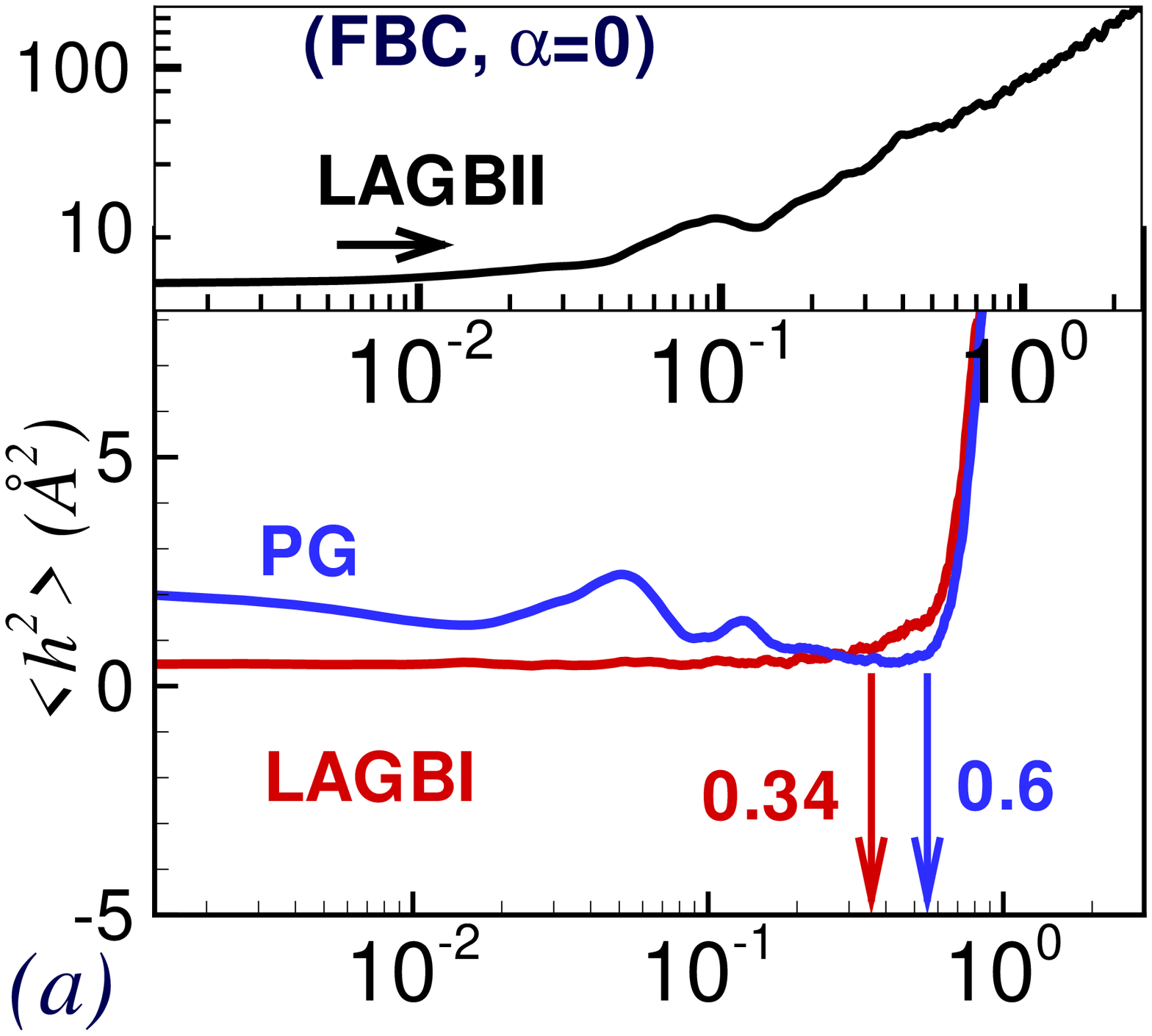}
\includegraphics[width=0.25\linewidth]{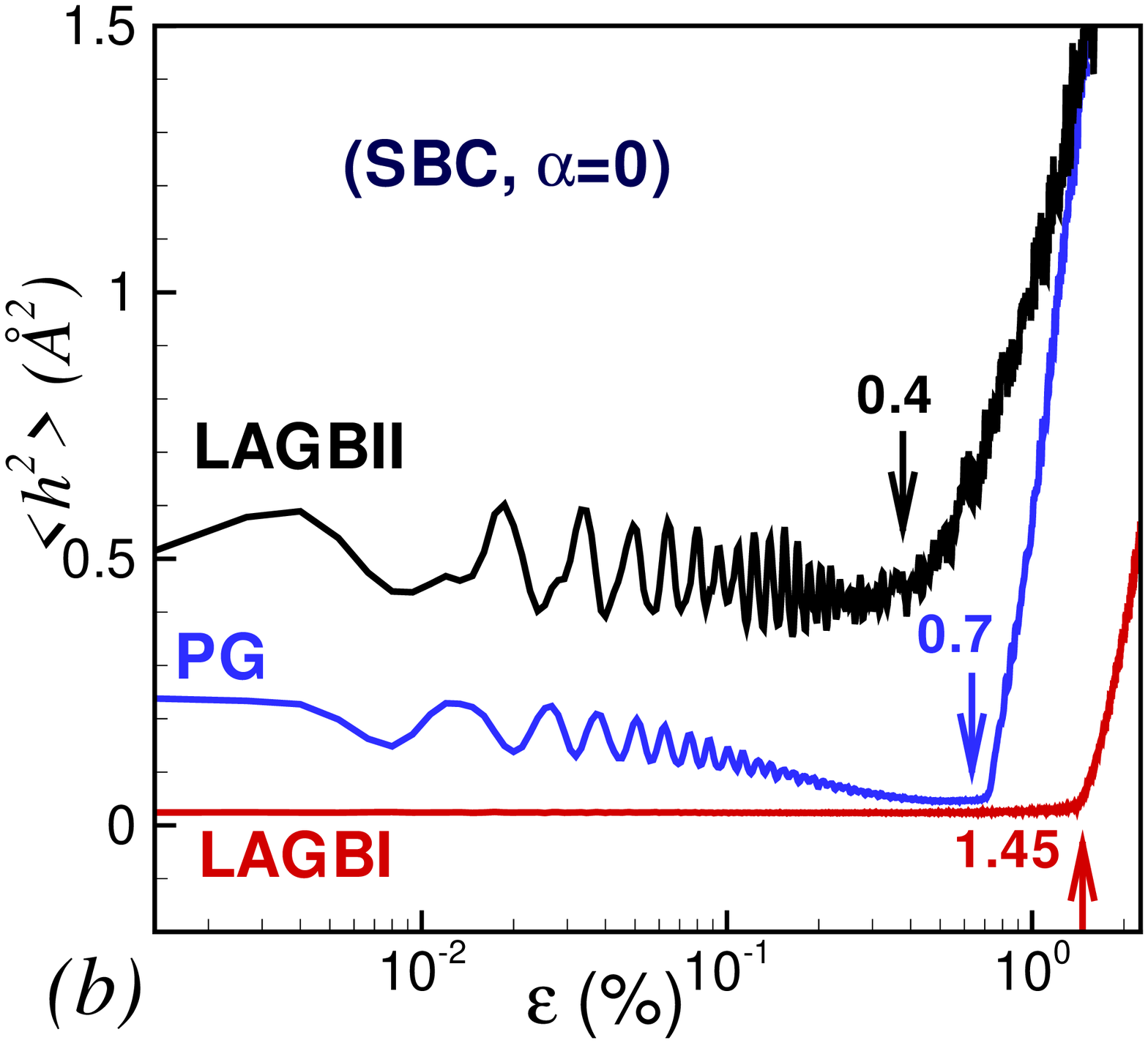}
\includegraphics[width=0.23\linewidth]{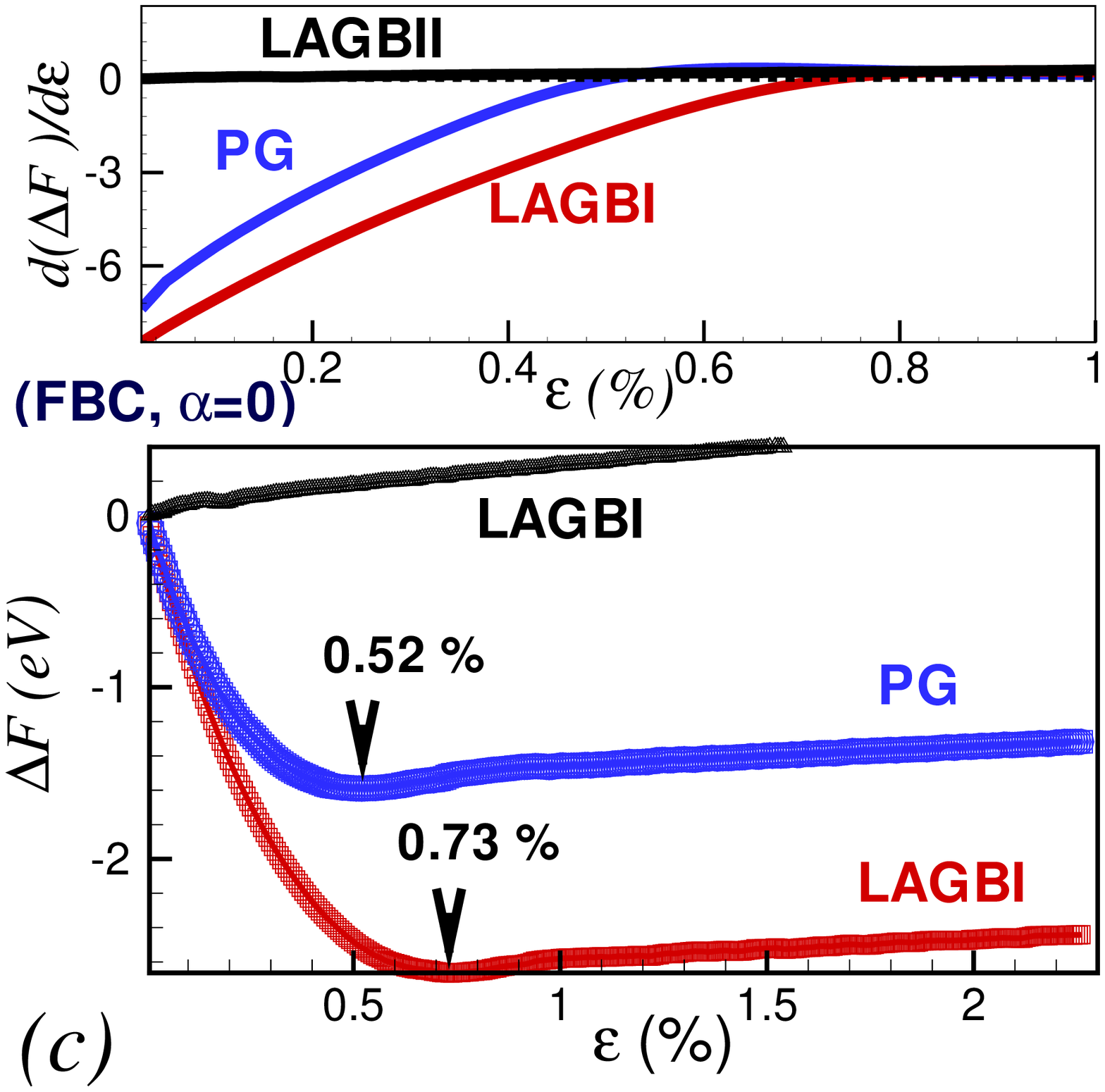}
\includegraphics[width=0.23\linewidth]{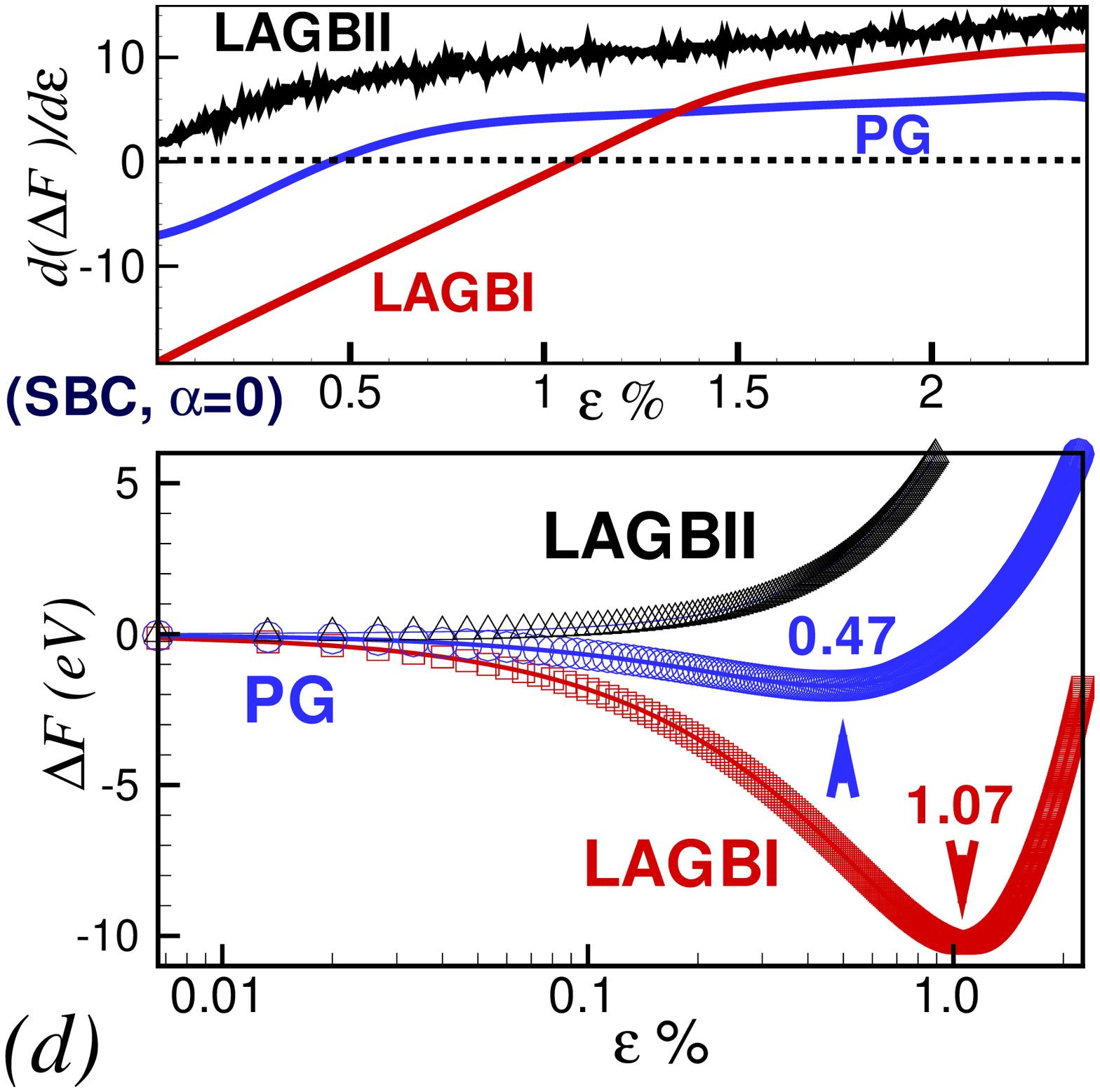}
\caption{(Color online) \textbf{(a) The out of plane displacement versus
applied strain  for graphene with free lateral boundary condition (a) and supported
 lateral boundary condition (b) with $\alpha=0$. Free energy change (bottom panels in (c,d))
 and its first derivative (top panels in
 (c,d)) during compression, for GNRs with (LAGBI, LAGBII) and without (PG) grain boundary
for FBC (c) and SBC (d).}
\label{figFree}}
\end{center}
\end{figure*}

%To calculate the change of the free energy of the GNRs, due to the
%application of an external force on the boundary, an equilibrium
%approach is no longer applicable and a non-equilibrium MD is needed.
We use the Jarzynski equality~\cite{jar}, i.e. $\Delta F=-\beta^{-1}\ln\langle{\exp(-\beta W)}\rangle$,
 which gives a relation between
the difference of the free energy and the total work done on the
system ($W$) during a non-equilibrium evolution where $\beta=1/k_BT$.
 The averaging is done over several  realizations  of
the paths between the initial and the final state.
We found that averaging over 10 simulations
with different initial states for each particular case, results in a
sufficient accurate value for $\Delta F$ (more technical details can be found in \cite{JPCMneek,PRLneek}).

Elasticity theory predicts that the shape of
the lowest mode of the  buckled state (of a simple bar with length $l$, under
axial symmetric load applied at its longitudinal ends and free from
lateral ends), is half a
 sine wave, i.e. $\delta w=\tilde{w}~\sin(\pi x/l)$ where $\delta w$ is the transverse deflection~\cite{book,prb82neek}.
  %Higher modes are possible only if the column is
%physically constrained from buckling into the lower modes by
%supporting mid points~\cite{book}.
For a rectangular plate subjected to the SBC, elasticity theory~\cite{prb82neek,book} predicts the
following possible deformations
\begin{equation}
\delta w=\sum^{\infty}_{m,n=1}\tilde{w}_{mn}\sin(n\pi x/a)\sin(m\pi
y/b), \label{dw}
\end{equation}
where $(m,n)$ are integers
in order to satisfy the SBC and
$\tilde{w}_{mn}$ is the amplitude of each mode $(m,n)$.
 Including the appropriate strain energy and using
Eq.~(\ref{dw}), the minimum buckling boundary stress
 for the considered systems always occurs for $m=1$
  and various values of $n$. It is equivalent
to a single half wave in the lateral direction and various harmonics
$n$ in the  direction of compression (i.e. perpendicular to the
grain boundary).

From our simulations we found that after many compression steps GNRs starts to
buckle, but the shape of the deformed GNRs was found to depend
strongly on the presence of the grain boundary and on the direction
of applied compression. Figures~\ref{fig2}(a,b,c) (FBC) and Figs.~\ref{fig2}(d,e,f)
(SBC) show snap shots of the deformed GNRs without (a,d)
and with grain boundary (b,c,e,f), beyond the buckling threshold, where $\epsilon=2.45\%$ and $\alpha=0$.
\textbf{The strain is calculated using $\epsilon=2\delta x/l$, where $l=a$ ($l=b$)
for $\alpha=0$ ($\alpha=\pi/2$).}
 From Figs.~\ref{fig2}(a,b,c) we see that the deformed shape for PG is similar to
 half a sine wave which is much less the case
for the LAGBI and LAGBII. The deformations in Figs.~\ref{fig2}(d,e,f) satisfy the condition $m$=1 in Eq.~(\ref{dw}),
while in the direction of the applied compression for LAGBI and LAGBII the shape of the deformation is different
from a sine wave which is most clearly seen around the grain boundary.

The buckling threshold, i.e. $\epsilon_b$, is measured by finding
 the sudden increase  in the average quadratic
out of plane displacement of the GNR atoms ($\langle h^2\rangle$).
The variation of $\langle h^2\rangle$, averaged over ten simulations
for $\alpha=0$, versus $\epsilon$ are shown in Figs.~\ref{figFree}(a,b) for FBC and SBC, respectively.
The vertical arrows indicate the transition points to the buckled state. The buckling strains, $\epsilon_b$,
are listed in Table~\ref{table2} for various situations.
\textbf{We found that graphene with LAGBII grain boundary subjected to FBC vibrates quickly
 so that the relaxed system (before compression) is buckled and thus the buckling strain is zero.
This is due to the larger angle  misorientation (32.2$^o$).}
However, notice that before buckling (FBC) $\langle h^2\rangle$ for PG fluctuates which is
less for LAGBI.
% which shows that in the ac direction ($\alpha=0$),
%the edges of PG have more dangling bonds as compared to the LAGBI.
Note that  a study of the edge reconstruction needs an ab-initio
approach which is out of the scope of the present study~\cite{edge}.

Changing $\alpha$ varies $\epsilon_b$ significantly.
In fact for $\alpha=\pi/2$, the buckling strain for  LAGBI is three times larger than for PG indicating
a considerable change in the structural deformation of graphene  when it is subjected to
compressive tension along the grain boundary.\textbf{ As seen from Fig.~\ref{figFree}(b),
for SBC the largest (smallest) buckling strain is for LAGBI (LAGBII)
and therefore we conclude that graphene with LAGBII
 is thermodynamically less stable as compared to LAGBI and PG. This is also confirmed by our
  free energy calculations.
  Note that a larger angle $\alpha$  grain boundary results in  weaker graphene~\cite{sciencegrain}. }

\begin{table}
\begin{tabular}{|c|ccc|ccc|}
\hline
$\alpha$&PG&LAGBI&LAGBII&PG&LAGBI&LAGBII\\

&FBC& FBC &FBC&SBC& SBC&SBC\\
\hline
0&0.6$\%$&0.34$\%$&0.01$\%$&0.7$\%$&1.45$\%$&0.4$\%$\\
$\pi/2$&1.07$\%$&3.0$\%$&0.2$\%$&-&-&-\\
\hline
\end{tabular}
  \centering
\caption{ Estimated  buckling strains for different boundary conditions and
different $\alpha$ with and without grain boundary.}\label{table2}
\end{table}

The obtained  buckling strains
are comparable  to the one obtained from recent buckling experiments i.e.
0.7$\%$~\cite{compressionamall,arxiv2010}. Our theoretical
buckling strains are a little smaller than those found in the
experiments, which we attribute to the presence of a weak
van der Waals interaction between the substrate and graphene in the experiment. Note that strains are more than
an order of magnitude smaller than those  where fracture occurs
in stretching simulations and nanoindentation experiments~(they were in the range 10-30$\%$)~\cite{lee}.

The change in the free energy difference when compressing the GNRs subjected to FBC (SBC)
 with $\alpha=0$ is shown in  the bottom panel of Fig.~\ref{figFree}(c) (Fig.~\ref{figFree}(d)).
 Notice that our non-compressed GNRs
(in the beginning of the simulations) are flat honeycomb lattice
structures (and during the first equilibration we did not change its
size) which are not in a thermomechanical equilibrium state at
finite temperature. Therefore, the free energy of this state should
be higher than the one of the equilibrium state. \textbf{The free energy of  graphene with LAGBII
increases with $\epsilon$, i.e. there are no minima in the free energy curves either for SBC or for FBC. This is a confirmation of our
previous argument about the instability of this system when it is suspended.}

%It is well known that at room
%temperature the equilibrium state of suspended graphene is not
%exactly a flat sheet and  intrinsic ripples are
%present~\cite{fasolinonature}. One of the major issues using
%molecular dynamics simulations is to find the relaxed state of the
%system. We found the minimum points in the free energy curve where
%the optimum strain is smaller than the buckling strain for both PG
%and LAGBI.  It implies that, at the optimum length (with optimum
%strain) our suspended GNRs are rippled (not buckled) and the system
%is in the thermomechanical equilibrium state. With other words in order to realize a
%perfect flat graphene layer we have to pull the systems with
%$\epsilon_m$.
Notice that the LAGBI system with FBC (SBC) exhibits a minimum, \textbf{i.e. equilibrium state corresponds to the
minimum points in the free energy curve,} for larger strains,
 $\epsilon_m=$0.73$\%$(1.07$\%$) as
compared to $\epsilon_m=$0.52$\%$(0.47$\%$) for PG. The reason is that the LAGBI system
with a flat surface is much farther from thermodynamical stability
than  the flat PG.
%Moreover it is interesting that all buckled GNRs subjected to FBC
% are more stable than the initial non-compressed GNRs and the minima of
%the free energy curves occur for strains smaller than the buckling
%strains.
The top panels in Figs.~\ref{figFree}(c) and (d) show the first derivative of
the free energy for FBC and SBC, respectively. Here the transition is continuous because of the
finite size of the simulated
GNR.

 As seen from the bottom panel of Fig.~\ref{figFree}(d) at the minimum point in the free energy curve
 the rippled state has a lower
free energy as compared to the initial non-compressed GNRs; i.e. $\Delta
F(LAGBI)=$-10.5 eV, $\Delta F(PG)=$-1.5 eV. Therefore, PG (LAGBI)
needs less (more) compression steps to reach its equilibrium size.

%\section{Conclusions}
In summary, we found that  deformations of graphene
nano-ribbons that are subject to in-plane axial boundary
stresses is different when the graphene sheet contains
a grain boundary. In the presence of a grain boundary  the GNR subjected to  compression
 parallel (perpendicular) to the grain boundary has a buckling strain that is largest, i.e. 3$\%$ (lowest i.e. 0.34$\%$)
 when the lateral edges are free. \textbf{Large angle grain boundaries  result in 
 smaller buckling strain
 and into an instability when graphene is suspended.}\\

 This work was supported by the Flemish Science Foundation (FWO-Vl) and the Belgian Science Policy~(IAP).

\end{document}